\documentstyle[11pt,aaspp4]{article}

\slugcomment{To appear in Astronomical Journal}

\lefthead{Keller, Bessell, Da Costa}
\righthead{WFPC2 imaging of young MC clusters}

\begin{document}

\title{WFPC2 Imaging of Young Clusters in the Magellanic Clouds\altaffilmark{1}}

\author{Stefan C. Keller, M. S. Bessell \& G. S. Da Costa}
\affil{Research School of Astronomy and Astrophysics, Institute of Advanced
Studies, The Australian National
University, Private Bag, Weston Creek P.O., ACT 2611, Australia.\\
Email: stefan, bessell, gdc@mso.anu.edu.au}
\altaffiltext{1}{Based on observations with the NASA/ESA {\em Hubble Space
Telescope}, obtained at the Space Telescope Science Institute, which is
operated by the Association of Universities for Research in Astronomy, Inc.,
(AURA), under NASA Contract NAS 5-26555.}

\begin{abstract}
We have carried out Wide Field Planetary Camera 2 F160BW, F555W and F656N
imaging of four young populous clusters: NGC 330 in the Small Magellanic Cloud
and NGC 1818, NGC 2004 and NGC 2100 in the Large Magellanic Cloud. We report
photometric results for these four clusters, including identification using
photometric colours of the cluster Be star population. We present theoretical
WFPC2 and broad band colours and bolometric corrections for LMC and SMC
metallicities. The use of the far-UV F160BW filter enables accurate
determination of the effective temperatures for stars in the vicinity of the
main sequence turnoff and on the unevolved main sequence.
\end{abstract}




%

\section{Introduction}
 
In this paper we present new UV, visual and H$\alpha$ photometry obtained with
the WFPC2 camera on board the HST, of NGC 330, NGC 1818, NGC 2004 and NGC
2100, four young populous clusters in the Magellanic Clouds with main sequence
turnoff masses in the range of 9-12 M$_{\odot}$. A key feature of the
evolution of 8-20 M$_{\odot}$ stars is the treatment of convection and the
presence and degree of the extension of the convective core. This paper
presents the observational data that will form the basis of a quantitative
investigation for the presence of convective core extension, and its
magnitude, as constrained by these clusters. The further analysis is to be
found in a later paper (Keller et al. \cite{evolution}).

The fundamental basis for our understanding of the evolution of stars of
various masses is derived from the comparison between the observed
colour-magnitude diagrams (CMDs) of star clusters and the predictions of
stellar evolution theory in the forms of evolutionary tracks and
isochrones. This interactive process has provided much insight into the
physics of stellar evolution. Even though the grounds of stellar evolution are
well understood, there remain a number of points of uncertainty that may
potentially impact not only on current estimates of evolutionary parameters of
stellar clusters, but also on the more sophisticated results that use such
parameters as their basis. One such point of contention is the physical
causation of the observed extension of the main sequence (MS) beyond that
predicted by no overshoot, non-rotating stellar evolutionary models.

Convective core overshoot is commonly proposed as the mechanism for the 
extension of the convective core. However, more recently the role of rotation 
has been recognised as offering a more natural way to bring about the same end 
result, i.e. increased internal mixing, (Maeder \cite{maeder98}, Langer \& Heger \cite{langer98} and Talon et al. \cite{talon97}). However with these models in their infancy, our discussion here is conducted within the convective core 
overshoot paradigm.

On the theoretical front, discussion of the efficiency of convective core
overshoot has been addressed by several authors with different results,
ranging from negligible to substantial (see e.g. Bressan \cite{bressan81}). In
the absence of hydrodynamical models the amount of mixing is only weakly
constrained by physical arguments. In order to ascertain the correct amount to
apply within stellar evolutionary models we must infer this amount from the
populations within star clusters.


The difficulty has been to find a sufficiently large, young and coeval
population in which to search for signs of overshooting. Galactic clusters of
comparable age to those of the present study contain small numbers of stars
and individually offer little insight. The clusters of the present study
contain upwards of 10$\times$ the mass of their galactic counterparts. Within
these clusters the numbers of stars are such that statistically meaningful
confrontations with evolutionary theory are possible.

The promising nature of the four clusters studied in the present work has lead
to several previous ground-based and IUE studies. These studies have revealed
several anomalies with evolutionary models which do not include
overshooting. Problematic features include: the observed and predicted
temperatures of the MS turnoffs differ by up to several thousand degrees; the
observed and predicted turnoff luminosities also disagree by as much as 1 mag
in $V$ (Caloi and Cassatella ~\cite{caloi95}, Caloi et al. ~\cite{caloi93});
the relative luminosity of the red giants and the turnoff are not consistent
with model predictions for a coeval population (Caloi and Cassatella
~\cite{caloi95}).

These features are indicative of some degree of convective overshoot within
the population. However, further conclusions from these studies are limited
since most have been restricted to the brightest members, the only stars for
which accurate effective temperatures were attainable. They extend to just
below the MS turnoff. The data we present here are a significant extension to
these previous studies. The superior resolution of the WFPC2 camera and far-UV
coverage has enabled us to extract accurate effective temperatures for a large
sample of stars up to 4 magnitudes in $V$ below the MS turnoff. This forms a
database suitable for quantitative investigation of convective core overshoot.

\section{Observations and Data Reduction}
The data presented in this paper were obtained by the HST using the WFPC2 on
June 1997. Exposures in F555W, F160BW and F656N were obtained. Table
\ref{obslog} details the exposures obtained. The F160BW filter
($\Delta\lambda$= 446\AA, $\lambda_{e}$ = 1491\AA) is a wideband filter with
negligible red-leak. The imaging and transmission properties of the F160BW
filter have been described by Watson et al. (\cite{watson94}). F656N is a
narrow band filter($\Delta\lambda$= 22\AA) centred on the H$\alpha$ line. This
filter was included to identify those stars showing H$\alpha$ emission, namely
Be stars, of which the four clusters discussed here are known to have large
populations (e.g. Keller et al. \cite{bestars}).

\placetable{obslog}

Biretta and Baggett (\cite{biretta}) have examined the noise characteristics
of the far-UV flats used in the standard data pipeline. They find that the
excessive noise in the F160BW flat field is a serious limitation on the F160BW
photometry. We have reprocessed the F160BW data through the standard data
pipeline (Holtzman et al. 1995a) for bias subtraction and flat
fielding. Following Biretta and Baggett we have used a F255W flat instead of
the F160BW. To account for the large-scale vignetting of the F160BW filter in
the WFs we form a vignetting correction by taking the ratio F160BW flat/F255W
flat then smoothing by a 20 pixel FWHM Gaussian function and then dividing
this image into the data flattened with the F255W flat.

In both F555W and F160BW multiple exposures were taken, the set of images were
combined and cleaned of cosmic rays within the IRAF package using the GCOMBINE
task. Care was taken when combining these frames that the central regions of
individual stellar profiles were not truncated. When combining, the gcombine
task forms the median (or the average in the case of two frames) of each pixel
value after rejecting those values that are deemed statistically
unreasonable. Such selection is made on the basis of the readnoise, gain and
sensitivity noise.
 
 In some circumstances a side product of the cleaning/rejection process is
that some stars show truncated intensity profiles. We suspect that this maybe
due to subpixel shifts of the centroid of the stellar image between images in
the combining process. When the PSF is undersampled, as is the case with the
WFPC2, it is possible for the central pixel values of a stellar image to
differ by several sigma from that expected from noise calculations between
images due to small subpixel shifts. These suspect values are rejected, in all
cases the lower pixel value is retained. This effect is pronounced amongst the
brighter stars. We find a sensitivity noise of 0.1 is optimal within the
GCOMBINE task for avoiding this effect and ensuring cosmic ray removal.
 
The fields are relatively crowded in both F555W and F160BW. We found that
photometry through a 3px radius (0.15\arcsec on PC, 0.3\arcsec on WF) aperture
was optimal. Appropriate aperture corrections were made to the 0.5\arcsec
standard adopted by Holtzman (\cite{holtzman95b}). The sky brightness was
determined from the median within a 5px wide annulus of inner radius 10px. The
measured FWHM in the F160BW images varies significantly in all cameras in a
radial manner; in its centre the FWHM is around 1.7px and at the edges
$\sim$1.9px with a marked ellipticity. The measured FWHM for the F555W filter
is around 1.6px (0.08\arcsec on PC, 0.16\arcsec on WF). The PSF in F555W does
not vary significantly across the field. Tests with a range of apertures have
reassured us that the variable PSF in F160BW does not introduce position
dependent variations in the extracted photometry through our chosen aperture.

F555W and F160BW magnitudes are reported in the conventional system based upon
the spectrum of Vega. The zeropoints used were taken from
STScI:WFPC2(\cite{STSci:WFPC2}). We have corrected our F160BW magnitudes for
the attenuation due to adhering contaminants on the external aperture
following the prescription of Holtzman et al (\cite{holtzman95b}) and the
contamination rate given for chips 1 and 3 in Whitmore
(\cite{whitmore97}). The WFPC2 is subject to a variation in charge transfer
efficiency across each detector. We have made a correction using a simple ramp
model as described in Holtzman et al. (\cite{holtzman95b}).

A degree of geometric distortion is present in the WFPC2 camera (details in
Whitmore \cite{whitmore97}). This is most severe in the F160BW filter. For
simple aperture photometry as preformed here these distortions are not
problematic, however to minimise the chance of incorrectly matching faint
stars between the F160BW and F555W images it was found to be necessary to
transform the coordinates of the F160BW frame to those of the F555W frame.
 
Tables~\ref{ngc330phot} -~\ref{ngc2100phot}, available from CDS, report the 
photometric results for the four clusters.

\section{Photometric Uncertainties and Completeness}
 
As noted above, flat fields are a major contributor to the uncertainties in
our photometry. Whitmore \& Heyer (\cite{whit97}) have found that in the
optical broad-band filters the uncertainties introduced into aperture
photometry of point sources due to flat fields are of the order of 1.5\% or
less. This is not the case in the far-UV, here Biretta and Baggett
(\cite{biretta}) report that the F160BW flat in particular is very noisy. The
RMS noise is found to be $\sim$20\% near the centre of the field. A most
demonstrative improvement offered by the adoption of the F255W flat is the
reduction by $\sim$1/3 in the apparent width of the upper MS in the resultant
CMDs.

We have compared the photometry in the short and long exposures for both F555W
and F160BW. This shows the internal accuracy of the photometry. We see that
the internal accuracy is of the order of 0.03 mag for the brightest stars,
ie. those with F555W$<$14.5, and rises to 0.1 for stars 4 mag. below the
saturation level of the long exposure (F555W=19.0 and F160BW=16.5). We find no
indication of a systematic dependence of the difference in magnitude between
short and long exposures with magnitude.

A number of photometric studies of these clusters exist in the
literature. Many of the earlier works confine themselves to the outer
extremities of the clusters. The first extensive study of these clusters is
the $B$$V$ photographic study of Robertson (\cite{robo}). CCD studies include
Walker (\cite{walker92}: NGC 330), Sebo \& Wood (\cite{seb94}: NGC 330),
Bencivenni et al. (\cite{ben91}: NGC 2004), Sagar et al. (\cite{Sagar91}: NGC
2004 and NGC 2100) and Balona \& Jerzykiewicz (\cite{bal92}: NGC 2004 and NGC
2100). However many CCD based studies have taken previous studies, in
particular that of Robertson (\cite{robo}), as the basis of
standardisation. Consequently there exist a limited number of external checks
on our F555W photometry. In the case of the F160BW magnitudes there exists no
straightforward means to quantify the uncertainties, we derive an estimate
from indirect means in section \ref{f160}.

\begin{figure}
\plotone{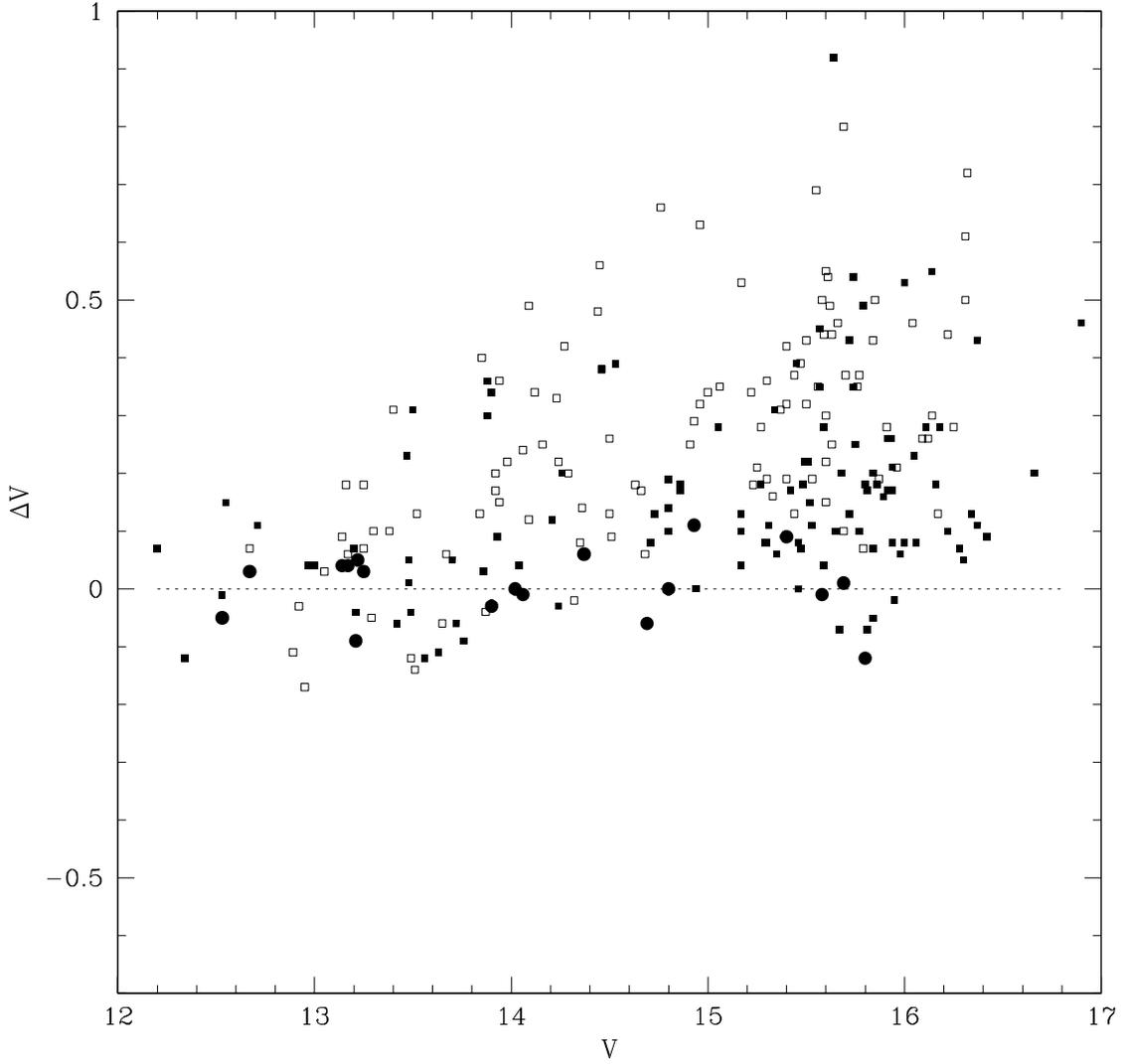}
\caption{Difference in $V$ magnitudes between our work and that of Robertson
(\cite{robo}) and Walker (\cite{walker92}) as a function of $V$ (i.e.
$V_{Rob,Walker}$$-$$V_{WFPC2}$). The filled boxes are those stars in regions
B,C or D in Robertson's nomenclature, the innermost A region stars are
represented by the open boxes. The filled circles represent the comparison
with Walker (\cite{walker92}).}
\label{comparison_of_mags}
\end{figure}

In deference to the frequent use of the magnitudes of Robertson in many
intervening discussions we firstly examined our photometric results relative
to those of Robertson.  For the purposes of comparison we have transformed our
F555W magnitudes to Johnson $V$ using the transformation of Holtzman et
al. (\cite{holtzman95b}). The results for all four clusters are shown in
figure~\ref{comparison_of_mags}. It is not suprising to see that there is a
much greater dispersion in Robertson's photometry of stars close to the
cluster core (open squares) where the degree of crowding is extreme, compared
to those more radially distant (filled squares). If we restrict the comparison
to those stars without close neighbours in Robertson's outer B,C,D regions and
$V$$<$14.0 we find a mean difference of +0.01$\pm$0.08 mag. The CCD-based
photometry of Walker (\cite{walker92}) in NGC 330 is of high precision. We
have included in figure~\ref{comparison_of_mags} the comparison between the
photometry of Walker and our own, here we see an agreement to -0.01$\pm$0.03
mag. This is quite satisfactory.

\placefigure{comparison_of_mags}

We have evaluated our completeness limits by the addition of sets of
artificial stars to the final frames for each colour. Using the IRAF ADDSTAR
routine we added 100 stars and observed the number recovered by the standard 
reduction procedure (using DAOFIND). This was repeated 100 times. An 
examination of the results finds our photometry $>$90\% complete to 19.5 in 
F555W, 17.5 in F160BW and 21.0 in F656N. This is consistent with the observed 
stellar luminosity function evident in figures~\ref{ngc330cmd} -~\ref{ngc2100cmd} 
which continues to rise to F555W$\sim$19, but drops sharply due to 
incompleteness beyond F555W=19.5

\section{Be Star Detection}

\begin{figure}
\plotone{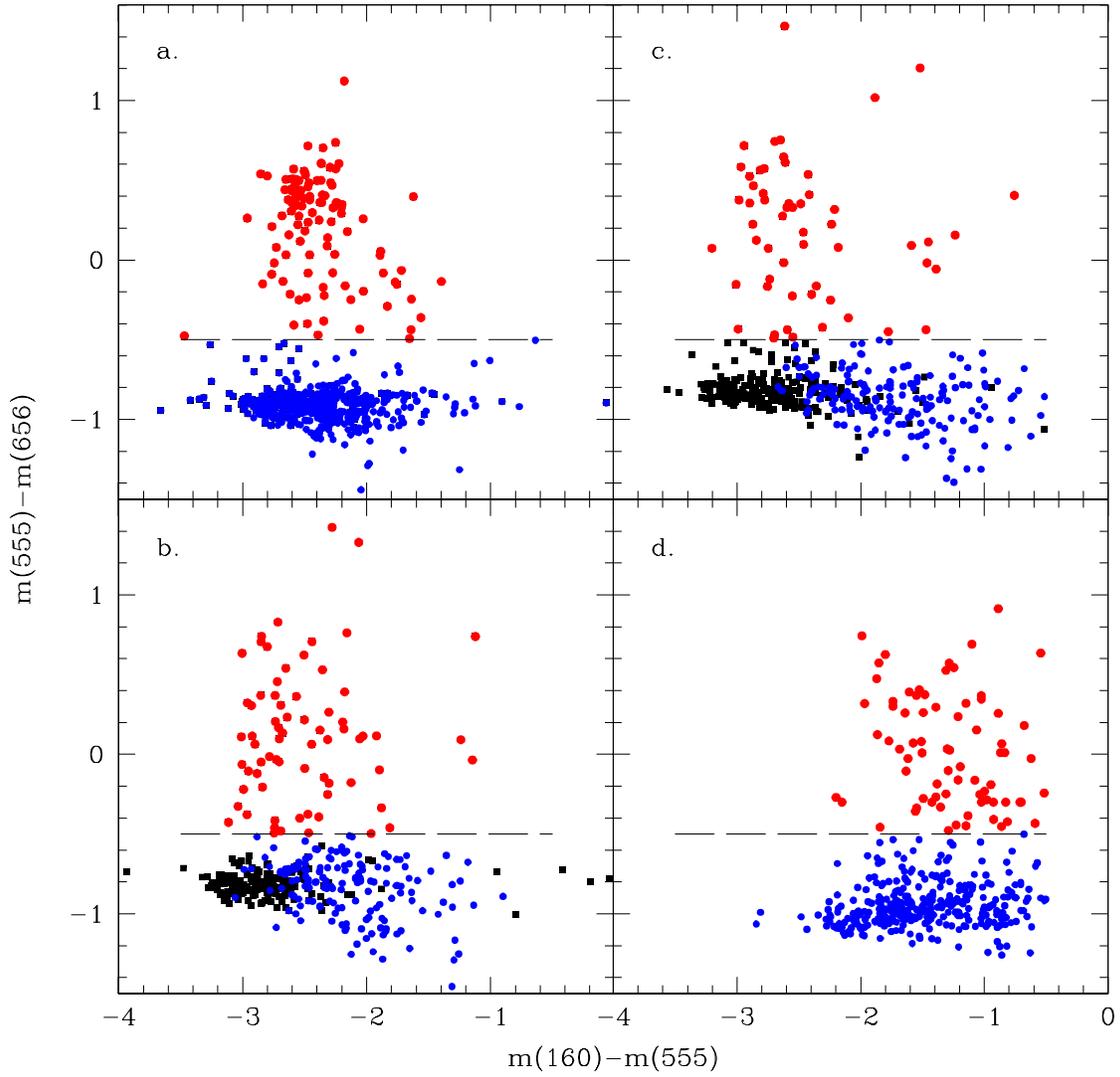}
\caption{The diagnostic colour-colour diagram for the identification of Be
stars in {\bf a.} NGC 330, {\bf b.} NGC 1818, {\bf c.} NGC 2004 and {\bf d.}
NGC 2100.  Be stars are identified as the stars above the dashed line (large
symbols).}
\label{bestars}
\end{figure}

The detection of the Be star population is made from the F555W-F656N,
F160BW-F555W diagram. Stars with strong H$\alpha$ emission should appear to
have a more positive F555W-F656N colour than their non-emission counterparts
at similar F160BW-F555W colour. Given that a typical Be star has H$\alpha$
emission with a full width at half maximum of $\sim$7\AA$\:$ and peak
H$\alpha$ flux 5 times the local continuum flux, a search of this kind with a
22\AA$\:$ filter is readily capable of detecting Be
stars. Figures~\ref{bestars}{\bf a-d.} show the diagnostic diagrams. Using
Fig.~\ref{bestars}{\bf a.} as a typical case, we note that most main sequence
stars lie in a tight, almost horizontal band. The cool supergiants are omitted
as they are saturated in H$\alpha$. Stars within this band do not have
detectable H$\alpha$ emission. Above this band are a group of stars that
clearly show significant H$\alpha$ emission: these are Be stars.

The dispersion in F555W-F656N of normal MS stars about the central locus of
the band is determined by the photometric errors in both the narrow band
H$\alpha$ and F555W magnitudes. Both have similar signal to noise
characteristics. This uncertainty will depend upon the magnitude of the
star. We limit our Be star selection process to stars with F555W$<$ 19.0. In
order to select a sample of Be stars, we draw a line in each of the
F555W-F656N diagrams parallel to the sequence of non-emission line stars and
at a distance of 0.5 magnitudes above it. The standard deviation in the
F555W-F656N colour of the normal B stars below this cutoff is at most 0.23 (in
the case of NGC 2100). Consequently the imposition of this cutoff excludes
members of the band of normal main-sequence stars to at least a 2$\sigma$
level.

Application of such a cutoff introduces a lower limit on the equivalent width
of the Be stars detected by our survey. We estimate that this limit is of the
order of that within Keller et al. (~\cite{bestars}) of
$\sim$9\AA. Spectroscopy in NGC 330 (Keller et al ~\cite{be330}) found, from a
sample of 29 B and Be stars, only one star with an emission equivalent width
smaller than this limit. For this reason we do not consider that our lower
limit is a significant limitation. All known Be stars within the outer
extremities of the clusters are retrieved by our Be detection criteria. Within
the cluster cores the high spatial resolution of the present data reveals a
large number of Be stars either undetectable or unreliably located within
ground-based data.

Nebular H$\alpha$ emission is visible in the field of NGC 2100. Implicit in the photometric detection technique is sky subtraction on small spatial scales which avoids mis-classification of normal B stars superimposed upon filamentary structure within the background unless the background is particularly clumpy which is not the case within our data.

\placefigure{bestars}

\section{Luminosity Function of the Be Fraction}

We have investigated the fraction of stars along the MS which are Be stars by
binning both emission-line and non-emission-lines stars with F160BW$-$F555W$<$
$-$1.0 into magnitude bins in F555W. Figure~\ref{bef}{\bf a-d.} show our
results. In each case the Be fraction peaks towards the MS
turnoff. Table~\ref{betab} examines the statistical significance of this trend
in more detail. Here we have divided the stars into two groups: the brighter
stars within 2 magnitudes of the MS turnoff (i.e. F555W$<$16.5) and the
fainter stars with 19$<$F555W$<$16.5, all with F160BW$-$F555W$<$$-$1.0. In
each case the fraction of Be stars near the MS turnoff are significantly
higher than the Be fraction further down the MS. We note that V=19 corresponds
to spectral type B9 in the calibration of Zorec and Briot (1991), consequently
below this cutoff we would not expect any Be stars.

\placefigure{bef}

\placetable{betab}

In Keller et al (\cite{bestars}) we performed a ground based survey for Be
stars within these clusters and their surrounding fields. Within this previous
study a similar analysis as that presented above, revealed no statistically
significant peak towards the MS turnoff. We note that the observations in
Keller et al (\cite{bestars}) were restricted by the imposition of a limiting
magnitude of $V$=17.5 (i.e. spectral type B5). In addition, due to crowding,
the inner 15\arcsec$\;$ of each cluster was unusable. The present sample is
also free from significant field contribution, which acts to dilute any
difference between the cluster and field populations.

Maeder et al. (\cite{mae99}) provide a review of previous studies of the Be
fraction within the MCs and various galactic clusters. Amongst the four
clusters studied here reference is made to the studies of Grebel
(\cite{grebel97}: NGC 1818), Grebel et al. (\cite{grebel92}: NGC 330) and
Kjeldsen \& Baade (\cite{kjeldsen}: NGC 2004). A comparison of the two sets of
results shows qualitative agreement in the case of NGC 330 and NGC 2004. In
the case of NGC 1818, Grebel (\cite{grebel97}) finds a Be fraction which is
consistently higher than our results and does not show signs of a drop off
with increasing magnitude. Towards the faint limit of the sample of Be stars
indicated by Grebel (\cite{grebel97}) we find a high proportion of objects
within crowded regions which do not show signs of strong emission. We posit
that the spurious detections in the data of Grebel arise from crowding within
the central regions.

Fractions as high as seen in the present study are unprecedented when seen in
the light of comparable studies of the Be fraction within the surrounding
field. Our examination of the field population of Be stars within the MCs
described in Keller et al. (\cite{bestars}) revealed the fraction of Be stars
to be more or less evenly distributed with magnitude at 15\%, in contrast to
the peaked distribution seen within the clusters. This is in line with the Be
fraction seen within the galactic field. Studies drawing a volume-limited
sample from within the galactic field face the difficulties of interstellar
reddening and depth effects, at best they limit the Be fraction to 5\% at B9
rising to around 20\% at B2 ($\rm{M}_{V}$=-3.5) (see figure 1 of Zorec and
Briot 1997). The evidence is suggestive that there is a fundamental difference
between the Be star populations in the clusters and the field which, if borne
out by closer scrutiny, would have important implications on the evolutionary
status of Be stars. This is however only suggestive at the moment, in a latter
paper (Keller et al. ~\cite{iau175}) we will attempt to resolve this issue
through the examination of a larger sample of the field population within the
MCs.

\section{Colour-Magnitude Diagrams}
\label{f160}
\begin{figure}
\plotone{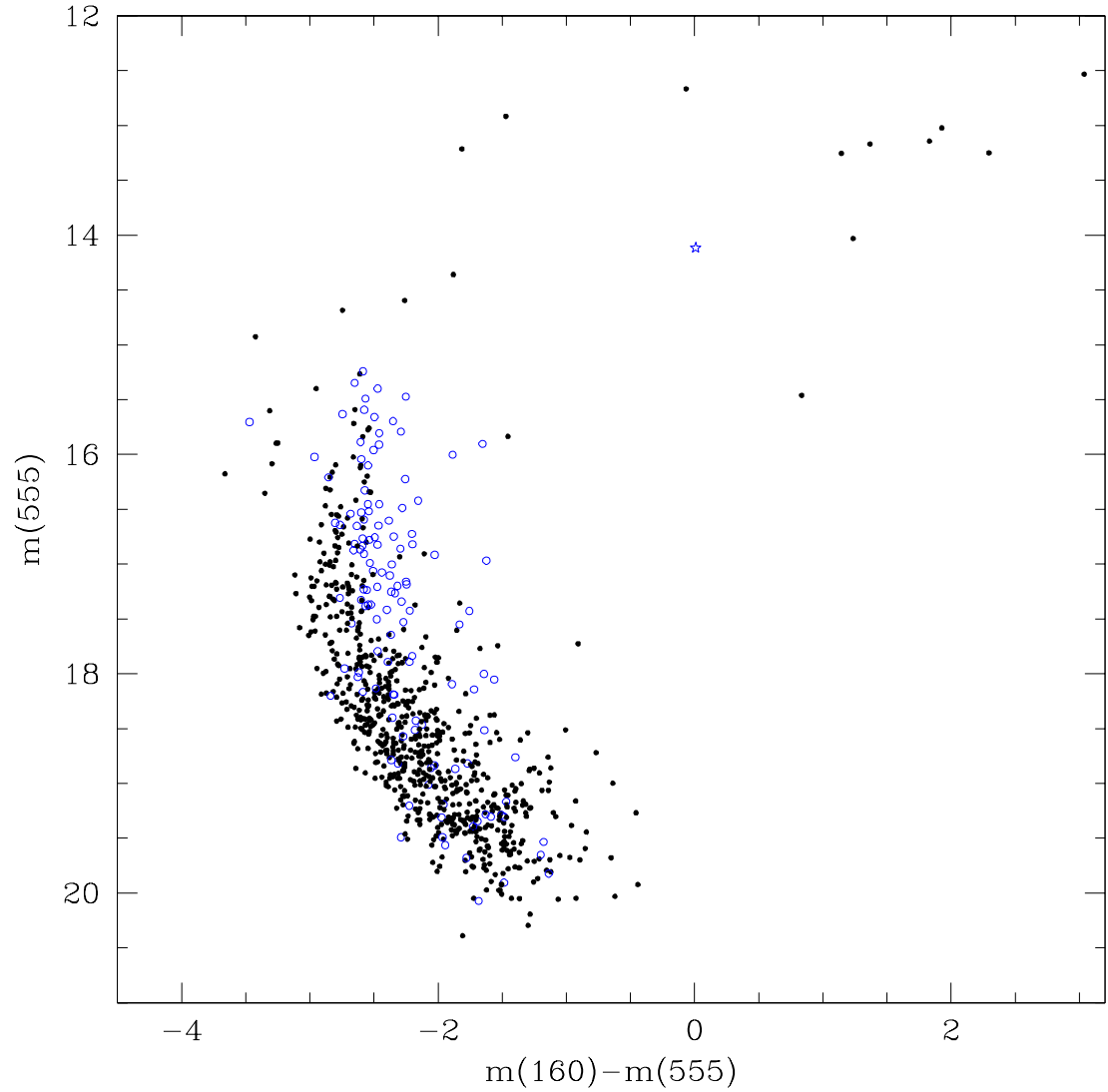}
\caption{Colour-Magnitude diagram for NGC 330. Be stars are indicated by the
open circles. The single star symbol at F555W$\sim$14 indicates the position
of A52, a system revealed by $J$,$K$ photometry to be comprised of a MS star
and a red supergiant (see text).}
\label{ngc330cmd}
\end{figure}

\begin{figure}
\plotone{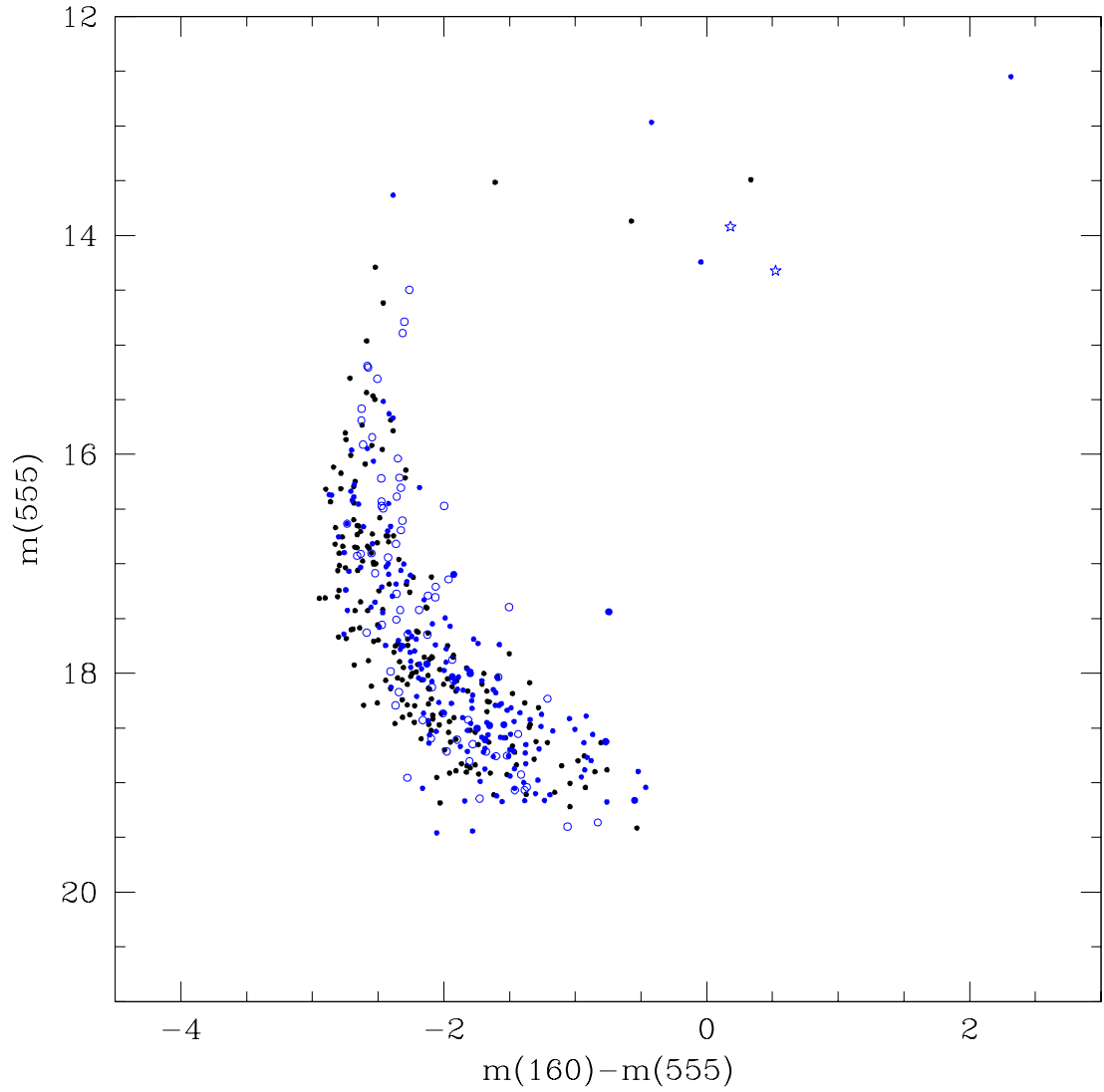}
\caption{Colour-Magnitude diagram for NGC 1818. The two star symbols at
F555W$\sim$14 indicate the positions of A18(left) and A95(right). Other
symbols as in figure~\ref{ngc330cmd}. }
\label{ngc1818cmd}
\end{figure}

\begin{figure}
\plotone{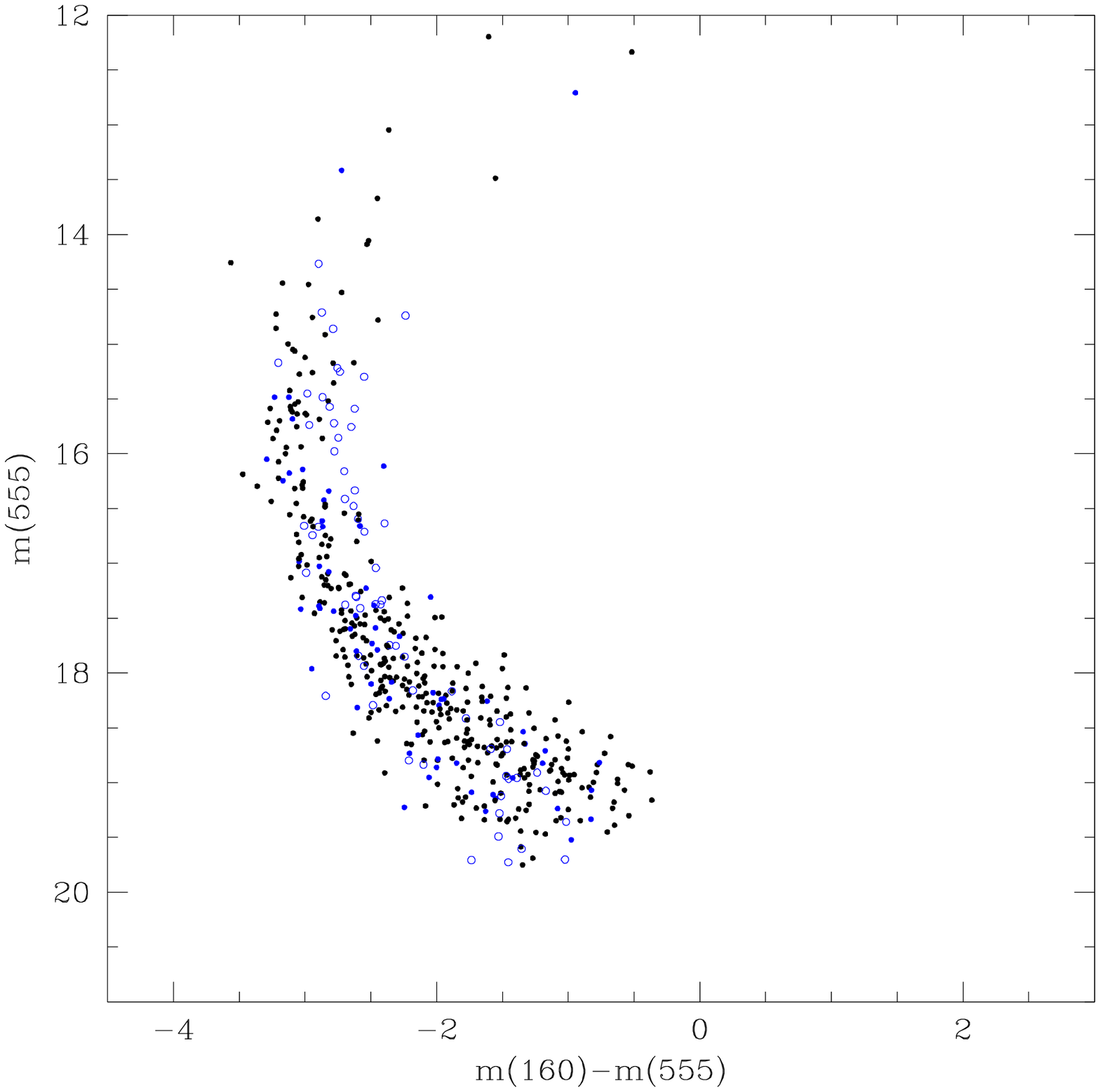}
\caption{Colour-Magnitude diagram for NGC 2004. Symbols as in 
figure~\ref{ngc330cmd}. }
\label{ngc2004cmd}
\end{figure}

\begin{figure}
\plotone{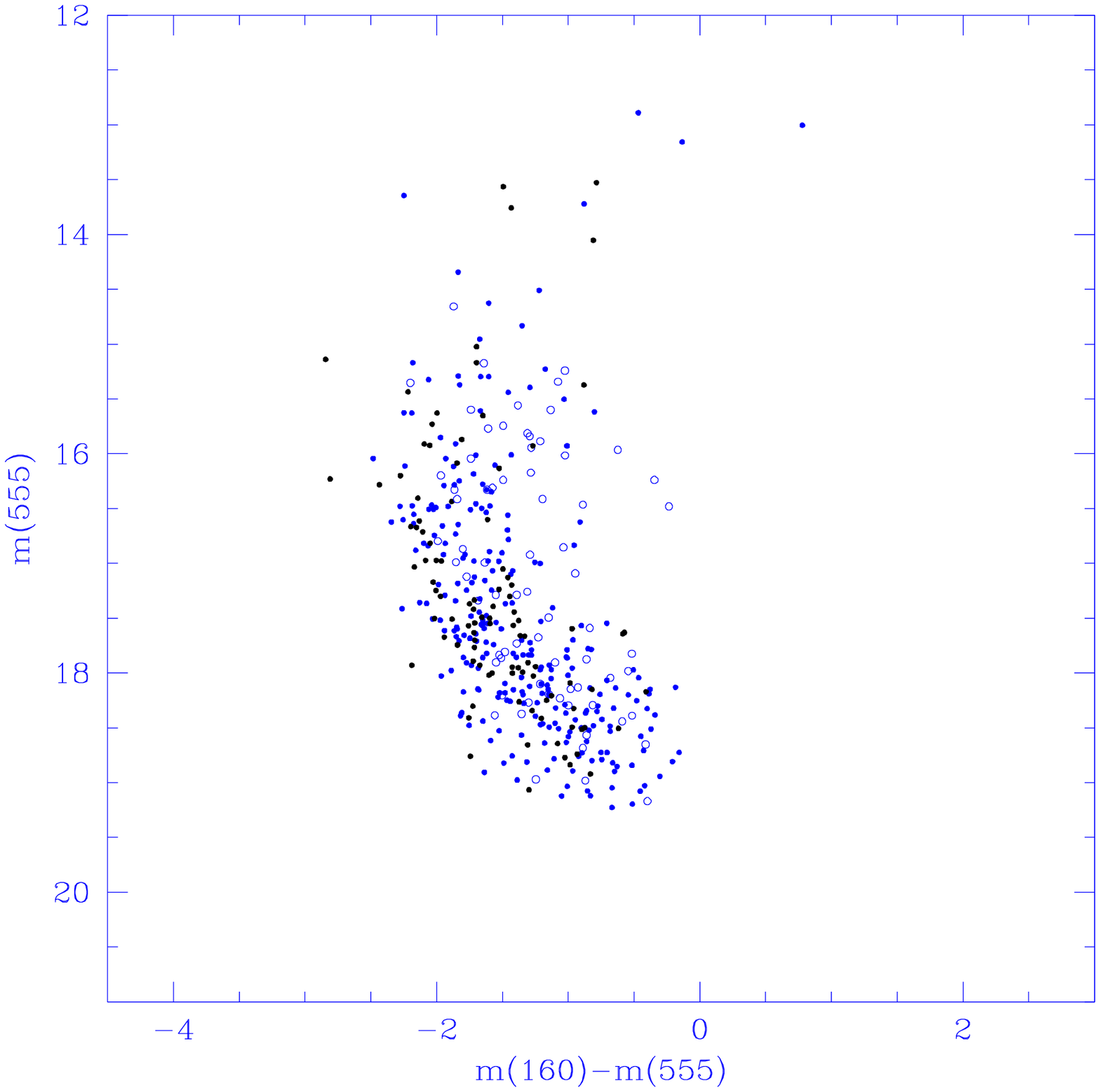}
\caption{Colour-Magnitude diagram for NGC 2100. Symbols as in
figure~\ref{ngc330cmd}. }
\label{ngc2100cmd}
\end{figure}

\begin{figure*}
\plotone{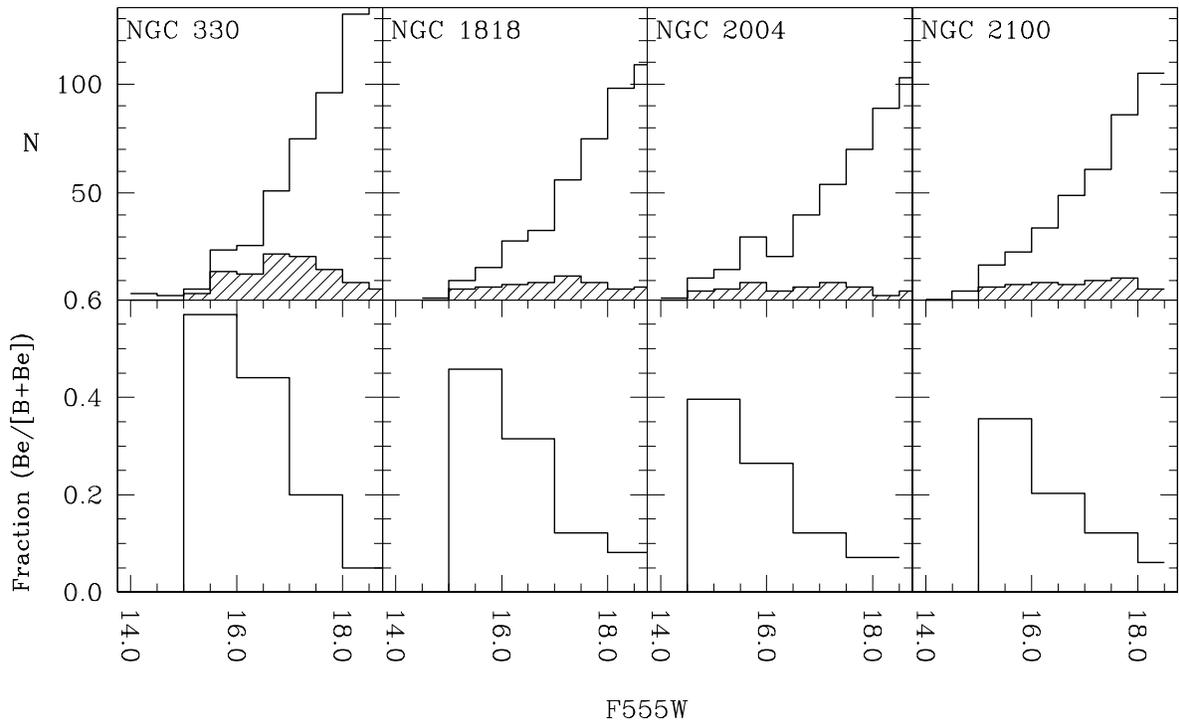}
\caption{{\bf Top.} Histogram of the number of Be stars (shaded) and
main-sequence stars (F160BW$-$F555W)$<-1.0$, including Be stars) in
half-magnitude bins down the main-sequence. {\bf Bottom.} The ratio of Be
stars to main-sequence stars in magnitude bins.}
\label{bef}
\end{figure*}

The colour-magnitude diagrams (CMDs) for NGC 330, 1818, 2004 and 2100 are
shown in figures~\ref{ngc330cmd} -~\ref{ngc2100cmd}. Figure~\ref{ngc330cmd}
demonstrates the most pertinent features. The MS is seen to terminate at
F555W=15.5. A prominent clump of A supergiants is seen around
F160BW$-$F555W$\sim$1.5. These are core He-burning blue supergiants. In the
case of NGC 2004 and NGC 2100 there is little sign of a clump of A
supergiants, rather the stars ``flow'' in a continuous manner from the tip of
the MS. Note the omission of the red supergiant population, which is
undetectable in F160BW. The CMDs show little sign of significant field
contamination.

Attention is drawn to the group of hot and luminous stars within NGC 330 which
are separated from the bulk of the cluster MS. The stars occupy a region bluer
than the MS (F160BW$-$F555W$>-$3.1 and 16.4$>$F555W$>$14.8). This region is
that typically associated with the position of blue stragglers in lower mass
clusters. Like blue stragglers in other clusters the central condensation of
this population is remarkable; of the 8 stars all but one (B13 discussed in
section 8) reside on the PC chip containing the cluster core (the
corresponding figures for the total population are: 407 stars on the PC chip
brighter than F555W$<$19.0 and 339 on the WF chips). We tentatively identify
these stars as blue stragglers.

The width of the main sequence is clearly greater than the estimated internal
uncertainties discussed above. The width likely results from the combination
of position dependent errors from the F255W flatfield (as discussed above),
differential reddening across the cluster and the intrinsic width of the main
sequence within each cluster, which is notably enhanced by the Be star
population. The precise contribution of each component is impossible for us to
separate. However we can gain an estimate of the relative importance of the
three major constituents through an examination of NGC 1818.

In the study of NGC 1818 by Hunter et al. \cite{hunter} presents photometry in
F336W, F555W and F814W. The F336W filter (analogous to Johnson $U$) is a
frequently used filter with well determined flatfield. Therefore let us assume
that the width of the MS exhibited in the F336W$-$F555W colour is due to
differential reddening and the intrinsic width of the MS. The width of the MS
in the photometry of Hunter et al. is $\sigma$=0.15mag (for stars
17$<$F555W$<$20). Using the colour excess ratios from Table~\ref{hsttab}
(discussed below) gives a corresponding $\sigma$=0.4mag in F160BW$-$F555W. We
see in Fig.~\ref{ngc330cmd} a width of $\sigma\sim$0.45mag for
16$<$F555W$<$18. We conclude from this that flatfield errors are a relatively
minor contributor to the dispersion evident in our CMDs. We consider a
$\sigma$=0.1mag an appropriate estimate for the uncertainty introduced by
flatfield errors in our quoted F160BW$-$F555W colours.

The effects of differential reddening is undoubtably the major contributor to
the large width of the MS in the case of NGC 2100. The cluster lies in the
vicinity of 30 Dor, a region filled with complex nebulosity. The width of the
upper main sequence is $\sim$0.6 and assuming 0.3 is due to intrinsic width of
the MS and flatfield noise this implies a $\sigma$E($B$$-$$V$)$\sim$$\pm$0.04;
a large but not unreasonable amount.

We have undertaken new IR photometry of these clusters described in detail in
Keller (\cite{irpaper}). The IR colours provide a valuable check on our WFPC2
colours for the brighter members. An examination of the $V$$-$$K$ and
F160BW$-$F555W colours has revealed a number of systems within the clusters
which consist of a binary pair of red supergiant and MS stars. These possess
F160BW$-$F555W $\sim$ 0-1, closely matched to those of the A supergiants, and
$V$$-$$K$ colours indicative of a red supergiant. The F160BW$-$F555W,
$V$$-$$K$ colours of these systems are the consequence of the combined light
in F555W but the light of only one member in F160BW and $K$, namely the blue
MS star in F160BW and the red supergiant in $K$. The systems highlighted as
such are: in NGC 330:A52 and NGC 1818:A18 and A95.

\placefigure{ngc330cmd}
\placefigure{ngc1818cmd}
\placefigure{ngc2004cmd}
\placefigure{ngc2100cmd}

\section{Transformation to the H-R Diagram}
\label{hrd}
Transformation of our measured F160BW$-$F555W colours and F555W magnitudes
into temperatures and luminosities was made through interpolation into a grid
of synthetically derived colours and bolometric corrections. Distance moduli
for the SMC and LMC were taken as 18.85 and 18.45 respectively, in line with
current determinations. The theoretical colours were computed for the revised
Kurucz (1993) fluxes used in Bessell, Castelli \& Plez (\cite{bessell98}) and
described in more detail by Castelli (\cite{castelli99}). The non-solar
metallicity is taken into account within these models ([Fe/H]=-0.5 LMC;
[Fe/H]=-1.0 SMC). In lower metallicity colours are bluer for a given
temperature. For example, a star of 20000K in both the SMC and LMC, appears
0.02 magnitudes bluer in the SMC.

For models cooler than 8000K the NOOVER grid of fluxes were used. The
passbands used were those in /cdbs/cdbs6/synphot\_tables/ and obtained via ftp
from ftp.stsci.edu. We computed the magnitudes for the F160BW, F336W, F439W,
F450W, F555W, F675W, F814W bands using a synthetic photometry program that
integrated the relative photon numbers for the various WFPC2 bands. The
zeropoints were adjusted to produce 0.04 mag for all bands with the Castelli
\& Kurucz (\cite{castellikurucz}) spectrum of Vega ($T_{eff}$/log g/[Fe/H]/ =
9550/3.95/-0.5). The UBVRI magnitudes were computed through energy integration
using the passbands of Bessell (\cite{bessell90}). These magnitudes were also
normalised to +0.04. Table~\ref{hsttab} (available from ADS) lists the grid of
colours and bolometric corrections.

\begin{figure}
\plotone{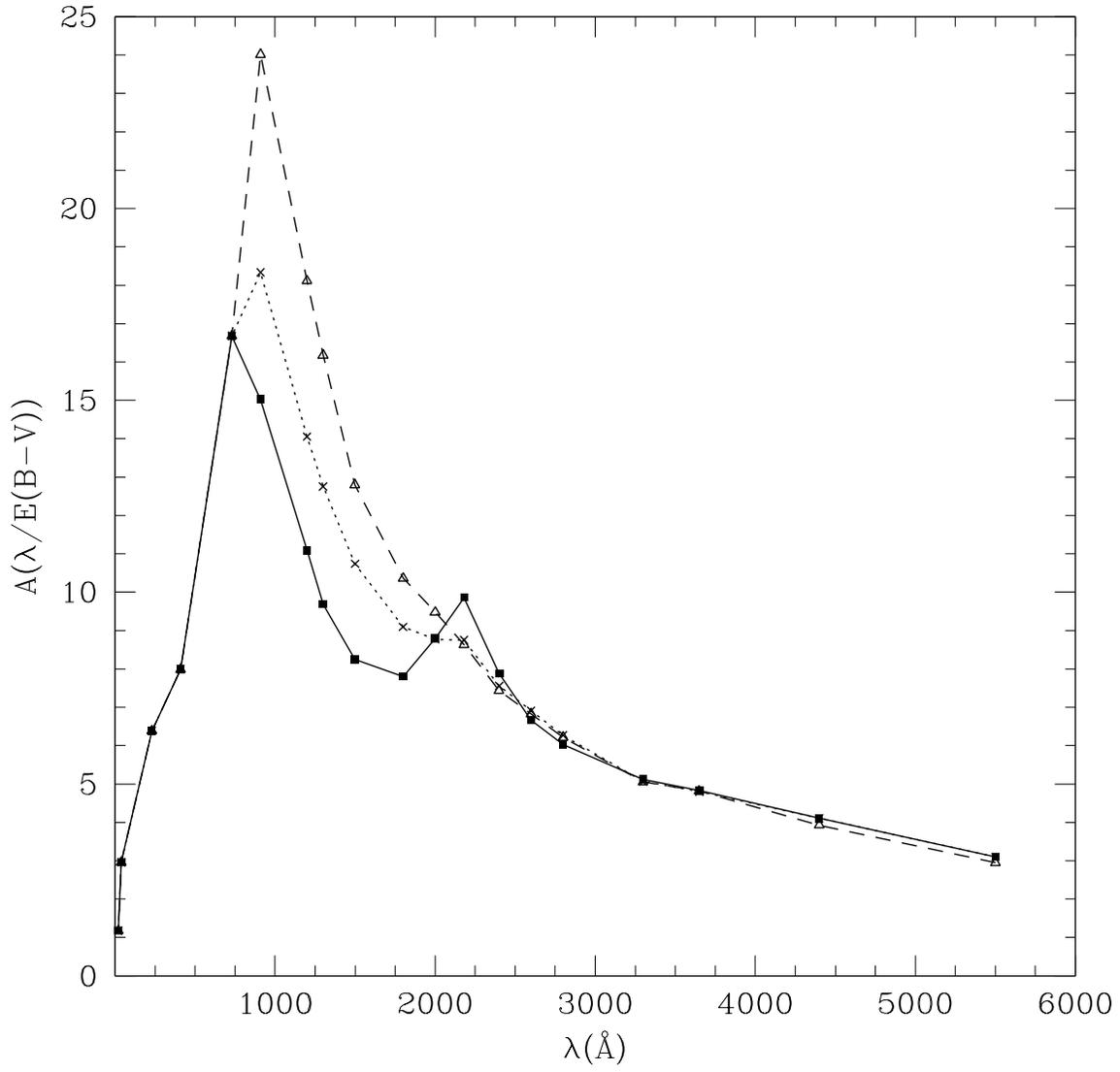}
\caption{The extinction curves adopted. The galactic component is described by 
the solid line, the LMC by the dotted line and the SMC by the dashed line.}
\label{extinction}
\end{figure}

To evaluate the effect of interstellar reddening we used the extinction curves
of Mathis (\cite{Mathis90}) for the Galaxy, Nandy et al. (\cite{Nandy81}) for
the LMC and Prevot et al. (\cite{Prevot84}) for the SMC. The LMC and SMC
curves were extrapolated to 90nm and set equal to the galactic curve for the
wavelengths redder than $V$. The galactic curve for diffuse dust (R($V$) =
A($V$)/E($B$$-$$V$)= 3.1) was used. We interpolated into the 22 magnitude
ratios A($\lambda$)/E($B$$-$$V$) of the extinction curves to produce
multiplicative factors at each of the 1221 Kurucz wavelengths for attenuating
the model fluxes. Figure~\ref{extinction} shows the extinction curve for the
optical and UV wavelengths. The colour excess ratio of E(F160BW$-$F555W) to
E($B$$-$$V$) under the different reddening regimes is given in
Table~\ref{colourexcess}.

\placetable{colourexcess}

\placefigure{extinctionplot}

The interstellar reddening to the four clusters was constrained by the
position of the MS in the CMD. The line of sight reddening was adjusted to
achieve a match between the essentially unevolved MS and that predicted by
standard evolutionary models. The reddening to both the LMC and the SMC were
considered in two components, half of the total reddening to galactic
absorption and the remainder due to intra-Cloud absorption. It was apparent
that the ``SMC'' extinction curve was not appropriate for NGC 330 as it
produced too large a F160BW extinction for the probable E($B$$-$$V$). We
adopted the LMC extinction curve for NGC 330. Table~\ref{reddening} reports
the values of reddening found in the present study. Uncertainties in these
values are of the order of $\pm$0.02 mag. The values are in agreement with
values in the literature.

\placetable{reddening}

\subsection{Effective Temperatures for the Be Star Population}
\label{beteff}
As can be seen in figures~\ref{ngc330cmd} -~\ref{ngc2100cmd} the Be population 
forms a secondary sequence at apparently cooler temperatures than the mean locus of non-emission stars in each cluster. The presence of a circumstellar envelope around Be stars is known to give rise to flux excess in various optical and infra-red bands. For this reason it is not clear that the observed colours and visual magnitudes are representative of the effective temperature and luminosities of the underlying star. 

We have sought an appropriate method for removing the effect of the
circumstellar reddening on the underlying star within the Be system. Numerous
studies have revealed that the flux emitted by a Be star in the $V$ band is
stronger than emitted by a similar B star without emission due to Balmer
continuum emission (Zorec \& Briot \cite{zorec91}; Kaiser \cite{kai89}). A
similar study by Zorec and Briot (\cite{zor85}) in the UV revealed no such
excess in UV flux amongst Be stars. This study made a comparison of
monochromatic magnitudes at $\lambda$=1460\AA$\:$ which is fortuitously close
to the central wavelength of the F160BW filter. This simplifies our task
considerably: we can consider $\Delta$(F160BW$-$F555W)=$-\Delta$F555W. Since
the $V$ excess arises from the circumstellar disk one would expect that the
$V$ excess would scale as the apparent surface area of the disk. In the same
way the strength of H$\alpha$ emission is similarly related to the apparent
surface area. Zorec and Briot (\cite{zor85}) show a clear correlation between
$V$ excess and H$\alpha$ emission strength.

We have used the details of this correlation and the F555W$-$F656N colours of
the Be stars within our sample to establish a $V$ excess for each object. In
order to do so it was first necessary to calibrate the F555W$-$F656N in terms
of the H$\alpha$ emission strength. This was achieved through the use of the
H$\alpha$ equivalent widths described in Keller et al. (\cite{be330}) obtained
three months following the present observations. Conceivably, during this
length of time the emission strength of some of the sample may have changed,
however this does not appear to seriously degrade the correlation between the
F555W$-$F656N colours and the H$\alpha$ emission strength.

The $V$ excess for each object from the above procedure has been applied to
F555W and to the F160BW$-$F555W colours for each object, making the Be stars
fainter and bluer. The figures~\ref{ngc330hrd} -~\ref{ngc2100hrd} show the
resulting HR diagrams, in which it is clear that the Be population has been
rectified, within observational uncertainties, to the locus of non-emission
line stars. This gives us considerable confidence in the corrections we have
applied to the Be stars within the CMD. These corrected colours are then
transformed to the H-R diagram as described in section~\ref{hrd}.

\section {Comparison with Previous Determinations of Effective Temperature}

\begin{figure}
\plotone{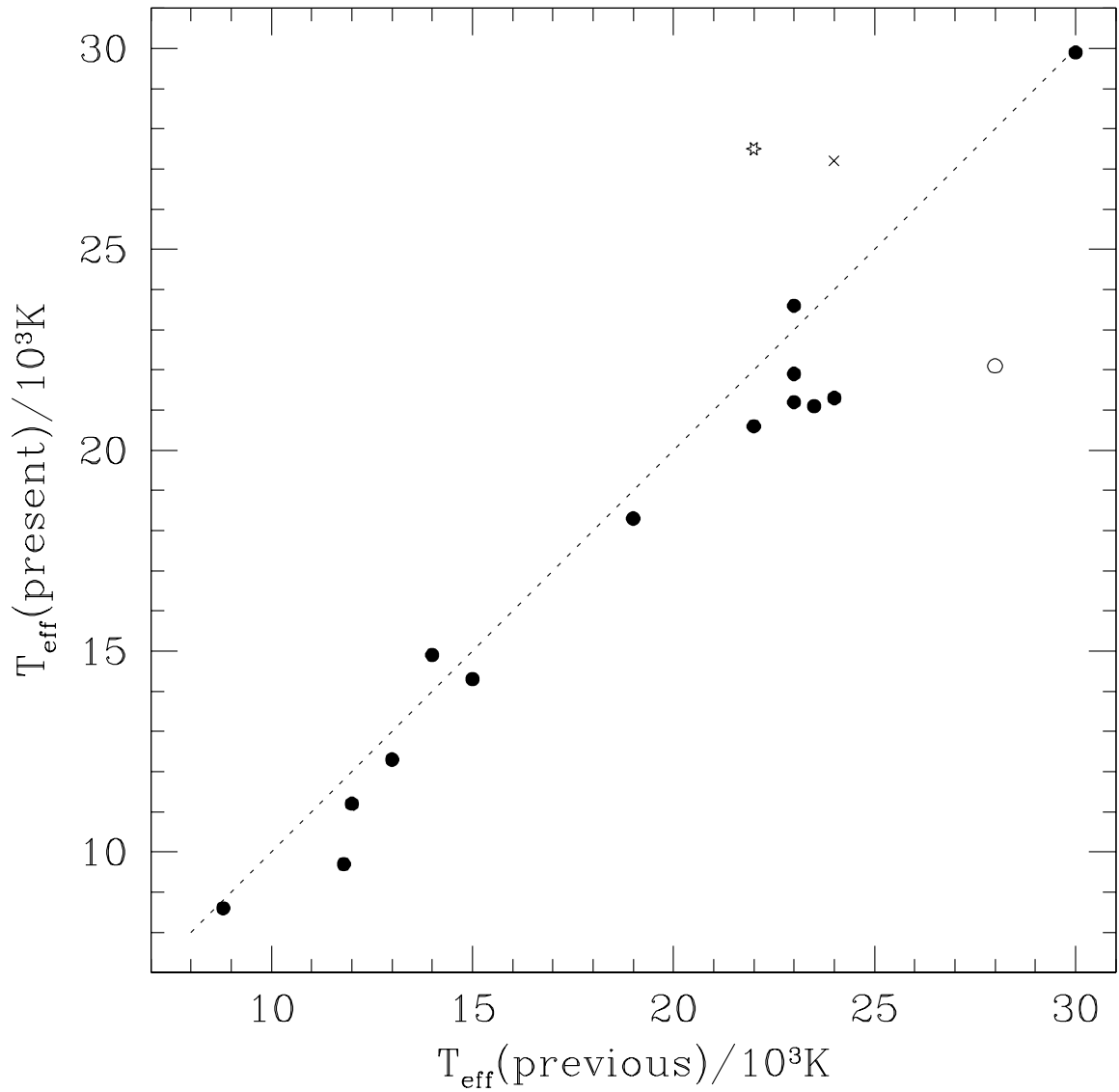}
\caption{Comparison of our effective temperatures with those of previous
studies (see text). The most discrepant points are NGC 330:B13 (star), NGC
330:A01 (open circle) and NGC 2004:D22 (cross).}
\label{comparison_of_temps}
\end{figure}

A number of studies have previously determined temperatures for a sample of MS
stars within these clusters. They have used a variety of techniques: the
studies of Caloi et al. (\cite{caloi95}; NGC 2004 and NGC 330) and
Bohm-Vitense et al (\cite{bohmvitense85}; NGC 2100) were based upon IUE
spectrophotometry while those of Lennon et al. (\cite{Lennon96}) and
Reitermann et al. (\cite{Reitermann90}) are from spectroscopic
analysis. Figure~\ref{comparison_of_temps} compares the $T_{eff}$ of these
previous studies with that of our own. Three discrepant points stand out, they
are A1 and B13 in NGC 330 and D22 in NGC 2004.

We have conducted spectroscopy of B13 and A1 using the Double Beam
Spectrograph on the 2.3m telescope at Siding Spring Observatory. Our spectra
consist of two simultaneously recorded segments, one blue (from 3400-4250\AA),
the other centred on H$\alpha$ (6200-6800\AA) at a resolution of
(0.6\AA/px). We spanned the Balmer discontinuity in our blue spectra to enable
an independent assessment of effective temperature.

B13 has a temperature of 27 500K from our photometry, making it spectral type
B0.5. In our HR diagram of NGC 330, B13 is placed in the region of blue
stragglers. Spectroscopically, however, this star has been designated by
Lennon et al. as B5III. The $B$$-$$V$ and $U$$-$$B$ colours of this system are
also discrepant; an observed $B$$-$$V$ of $-$0.19 (Walker 1992) indicates a
temperature of 30000K using the previously specified reddening, whereas the
$U$$-$$B$ colour ($-$0.85) gives 22000K. Our spectrum of this object shows a
narrow, weak H$\alpha$ emission (equiv. width = 9\AA, FWHM = 1.6\AA) which is
markedly different from the majority of Be stars which in general exhibit much
stronger and broader H$\alpha$ lines. Lines of Si-IV are strong and the Balmer
discontinuity is very small suggesting that the object is of spectral type
B1.5-B0 in reasonable agreement with our photometric temperature. The H Balmer
lines blueward of H$\beta$ show line cores that are unusually broad and
flat. This is perhaps an indication that the object is an approximately equal
mass binary system within an extended region of H$\alpha$ emission. Further
observations of this object with better signal to noise are required.

A1 has been designated B0.5III by Lennon et al. and Grebel et al. Our
observations make it much cooler at 22 100K (i.e. B2; Bessell et
al. \cite{bessell98}). Spectroscopically A1 has a Balmer discontinuity too
large and a Ca-II K feature too strong for B0.5III, rather we consider it to
be B2III, in line with our derived temperature. Existing photometric colours
have little to add, $B$$-$$V$ gives a $T_{eff}$ of 30 000K whilst $U$$-$$B$
gives 22 000K.

The third most discrepant point is that of D22 in NGC 2004. In the study of
Caloi and Cassatella, the determination of the temperature of D22 in NGC 2004 
was recognised as problematic (see discussion therein) due to the apparently 
discrepant $B$$-$$V$ colour.

If we leave aside these three most discrepant points we are left with a
scatter which we consider represents the relative accuracy of temperature
determinations drawn from optical and IUE spectroscopy. It is also apparent
that there is a minor systematic offset of our temperatures relative to that
of previous studies, in the sense that our temperatures are slightly
cooler. This amounts to 1500K at 25000K (i.e. 6\%) which is unlikely to be
significant given the uncertainties in previous measurements at such
temperatures.

\placefigure{comparison_of_temps}

\section{H-R Diagrams}

\begin{figure}
\plotone{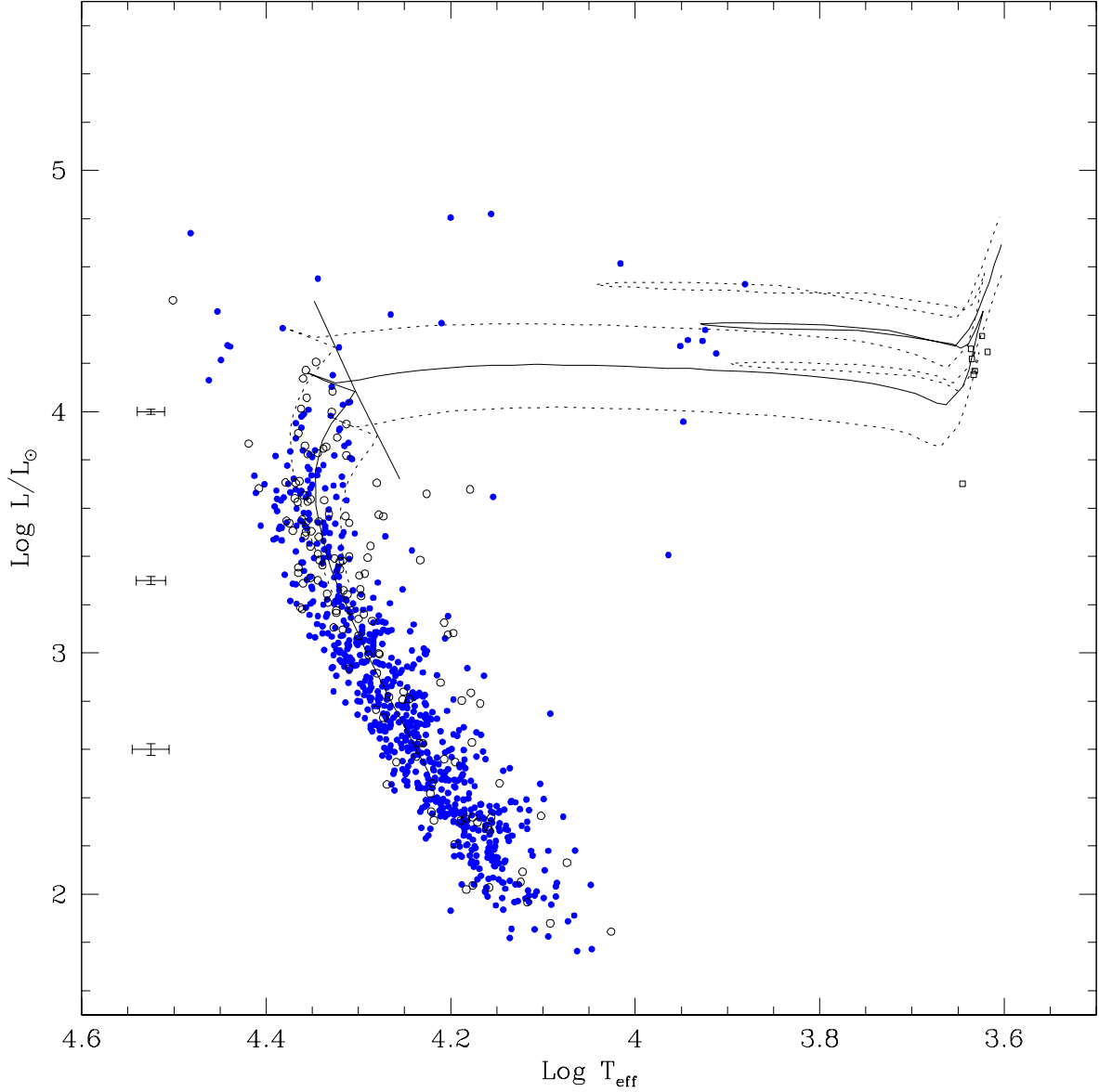}
\caption{H-R diagram for NGC 330. The location of Be stars after the
modification discussed in section~\ref{beteff} are indicated by open
circles. The red supergiant population within the WFPC2 field are represented
by the open boxes these are taken from Keller et
al. (\cite{irpaper}). Representative error bars for points {\it on} the MS are
shown. Indicative isochrones (log age=7.5 solid,7.6 and 7.4 dotted) are
overlaid (Fagotto et al.~\cite{fbbc} Z=0.004, overshoot) together with the
locus of the main sequence terminus (solid line).}
\label{ngc330hrd}
\end{figure}

\begin{figure}
\plotone{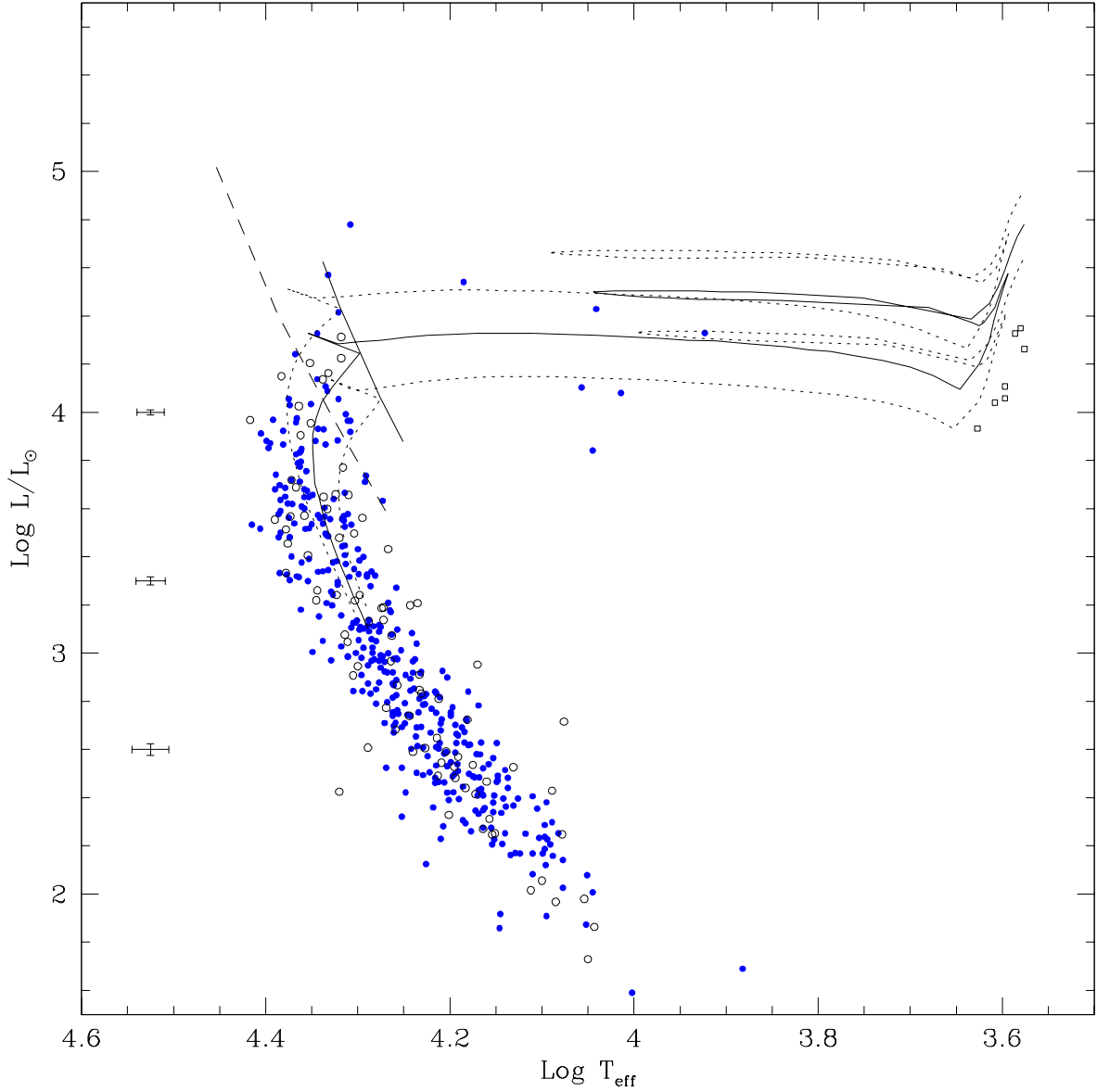}
\caption{H-R diagram for NGC 1818. Symbols as in
figure~\ref{ngc330hrd}. Isochrones are for log age=7.4 solid, 7.3 and 7.5
dotted. The locus of the main sequence terminus (solid line) is shown from
Bertelli et al. (\cite{bert94}, Z=0.008). The dashed line indicates the MS
terminus from models incorporating no overshoot (Alongi et al.~\cite{padova}).}
\label{ngc1818hrd}
\end{figure}

\begin{figure}
\plotone{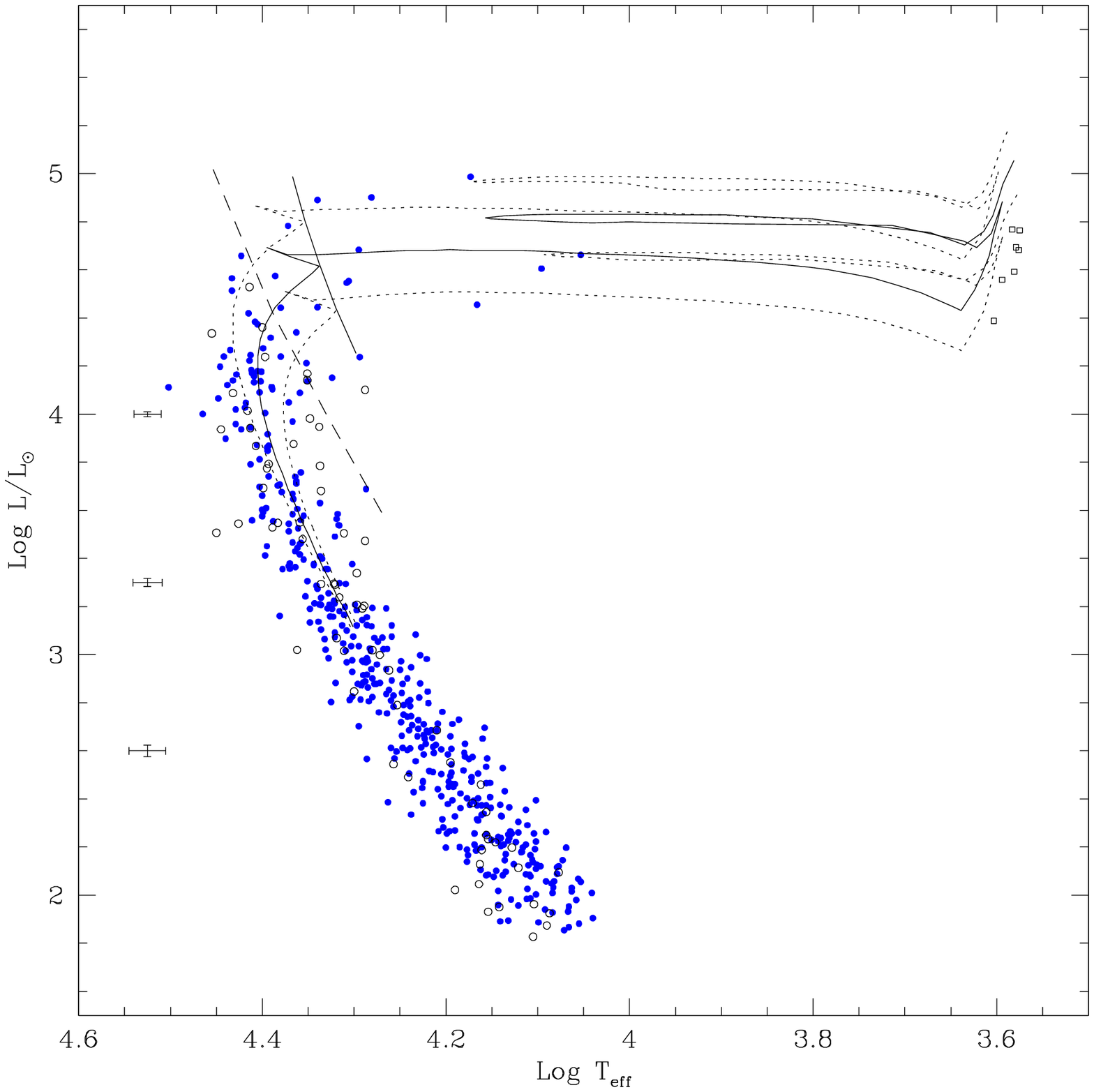}
\caption{H-R diagram for NGC 2004. Symbols as in
figure~\ref{ngc1818hrd}. Isochrones are for log age=7.2 solid, 7.1 and 7.3
dotted.}
\label{ngc2004hrd}
\end{figure}

\begin{figure}
\plotone{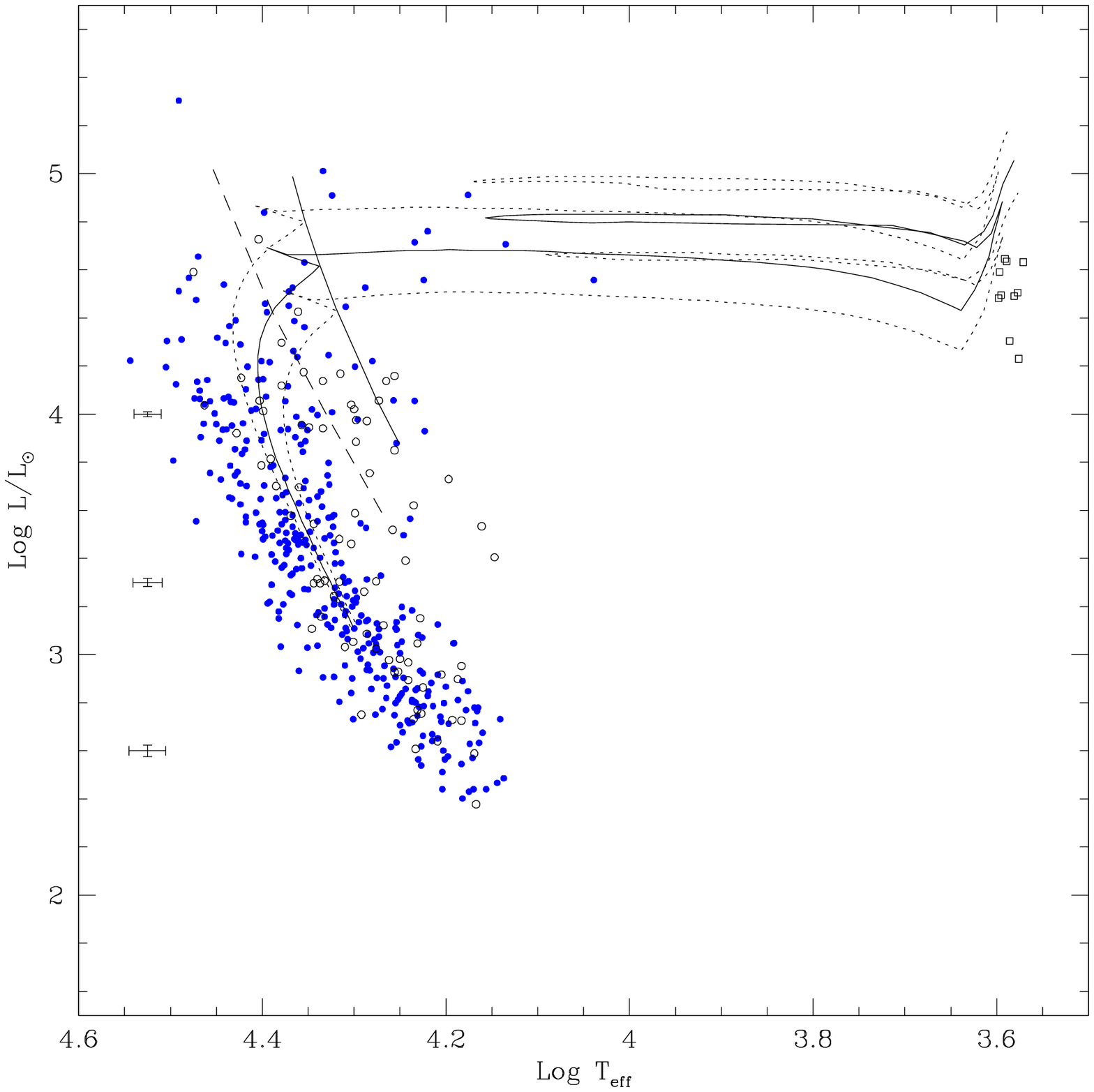}
\caption{H-R diagram for NGC 2100. Symbols as in
figure~\ref{ngc1818hrd}. Isochrones are for log age=7.2 solid, 7.1 and 7.3
dotted.}
\label{ngc2100hrd}
\end{figure}

The resultant H-R diagrams are shown in figures~\ref{ngc330hrd}
-~\ref{ngc2100hrd}. Also shown in figure~\ref{ngc330hrd} are typical error
bars for data points {\it on} the MS. The red supergiant (RSG) population
shown in these figures is that contained within the WFPC2 field. Effective
temperatures and luminosities for the RSG are taken from Keller et al
(\cite{irpaper}). We briefly discuss some of the implications of the H-R
diagrams here with reference to previous observations; detailed discussion is
deferred to Keller et al. (\cite{evolution}).

Overlain are isochrones from Bertelli et al. (\cite{bert94}) at 0.1 dex
spacing. Isochrones for Z=0.008 are used for the LMC clusters and Z=0.004 for
NGC 330. The discrepancy in the temperature of the RSG is discussed in Keller
et al (\cite{irpaper}). The ages of the clusters indicated by these isochrones
are of order log age=7.2 for NGC 2004 and NGC 2100, 7.4 for NGC 1818 and 7.5
for NGC 330. These ages are within the range of ages determined in previous
works.

In their study of NGC 330, Chiosi et al (\cite{choisi95}), claim that the
cluster data does not permit discrimination between convective criteria. They
are unable to fit the cluster CMD with a single age and require instead a
large spread of ages within the cluster. To a large degree this can be put
down to the insensitivity of the ($B$$-$$V$) colour to the temperatures in the
vicinity of the MS turnoff. The presence of Be stars, which have discrepant
$B$$-$$V$ colours, in the vicinity of the MS turnoff distracts further from
the clarity of the turnoff. This makes for a broad turnoff, which enables a
range of model isochrones to fit. Our data is a significant improvement upon
this. We also note that the tight grouping of the RSGs in these diagrams is
inconsistent with a large age spread.

\placefigure{ngc330hrd}
\placefigure{ngc1818hrd}
\placefigure{ngc2004hrd}
\placefigure{ngc2100hrd}
  
Within NGC 2004, Caloi and Cassatella (\cite{caloi95}) find that the
luminosities of the upper MS stars are inconsistent with these being
progenitors of the RSGs observed within the cluster. The latter concern is
resolved by the IR photometry of Keller et al. (\cite{irpaper}), which finds
the temperatures of these red supergiants to be of the order of 400K cooler
than previously determined. The consequent change in bolometric correction has
brought the luminosity of the RSGs into line with the top of the MS.

In figures~\ref{ngc330hrd} -~\ref{ngc2100hrd} we show the redward boundary of
the MS for standard ``moderate'' overshoot models (solid line - this is the
stage B of Bertelli et al. \cite{bert94}) and in figures~\ref{ngc1818hrd}
-~\ref{ngc2100hrd} models without convective core overshoot (Alongi et al
\cite{padova}). The region redward of this boundary and bluer than the evolved
A supergiants is a region which according to evolutionary models is traversed
rapidly. This region, the Blue Hertzsprung Gap (BHG), is expected to be devoid
of stars. It is clear from the HRDs that the no overshoot models fail to match
the density in this region. The predictions of standard overshoot models
provide a closer match. This necessity for a degree of overshoot is in
agreement with previous IUE studies of Caloi et al. (\cite{caloi93}) and Caloi
and Cassatella (\cite{caloi95}).

Whilst the standard overshoot models provide a closer match to the
observations a number of stars are seen in the four HRDs, which remain within
the BHG. The duration of evolution from the red edge of the MS to temperatures
cooler than the tip of the blue loop tracks is very rapid, it accounts for
0.07\% of the total lifetime for a 12$\rm M_{\odot}$ star (Fagotto et
al. \cite{fbbc}). Consider the HRD of NGC 2004. Here we see six stars clearly
within the BHG. The time spent in the BSG+RSG phases for such a star is 78
times longer than the traversal of the BHG. With thirteen BSG+RSG present we
would expect on this basis to find 0.2 stars present within the BHG.

The existence of populations within the BHG has been noted previously, the
controversy over their evolutionary status remains. Grebel et al
(\cite{grebel96}) in their study of NGC 330 have suggested that the population
of stars within the BHG is a mixture of rapidly rotating B/Be stars and blue
stragglers. In the light of our accurate determination of effective
temperatures we can rule out the possibility that these stars are blue
stragglers. Neither are any of the interloping stars in this region Be stars
from our study or at several other observational epochs (see Keller et
al. \cite{bestars}). Four of these BHG interlopers, A01, B22 and B30 in NGC
330 and D12 in NGC 1818, have been the focus of spectroscopic studies
(Reitermann et al \cite{Reitermann90}, Lennon et al. \cite{Lennon96}, Korn et
al. ~\cite{kor99}). These stars appear as normal B type stars, albeit some
with evidence of a N overabundance.

As discussed above, the temperature and luminosity of the terminus of the MS
is particularly sensitive to the degree of extension of the convective
core. Perhaps the stars within the BHG are an extension of MS evolution to the
red brought about by a degree of internal mixing in excess of that prescribed
in standard overshoot models. A more detailed analysis is required in this
regard and will be presented in our subsequent paper.

\section{Summary} 
Our WFPC2 photometry of NGC 330, 1818, 2004 and 2100 has shown that these
clusters form an excellent testing ground in which to examine a number of
outstanding issues in stellar evolution. The unprecedented resolution offered
by the WFPC2 camera provides us with a sample of sufficient size to enable a
statistically meaningful confrontation with standard evolutionary models. The
far-UV coverage has provided good temperature estimates for the hot main
sequence population. With this information it should be possible to ascertain
the presence and amount of internal mixing due to convective core overshoot
expressed in the population with the present data.

\begin{acknowledgements} 
SCK acknowledges the support of an APA scholarship and a grant from the DIST 
Hubble Space Telescope Research Fund. 
\end{acknowledgements}

\clearpage
\begin{table}
\begin{center}
\begin{tabular}{ccccc} \hline
Cluster Name & File Number  & filter   & exp. time &  \\ \hline
NGC 330 & U3JS0101M & F555W  & 1.6s\\
NGC 330 & U3JS0102M &  \dots      & 10s\\
NGC 330 & U3JS0103M &  F160BW     & 100s\\
NGC 330 & U3JS0104M &  \dots      & 800s\\
NGC 330 & U3JS0105M &  \dots      & 800s\\
NGC 330 & U3JS0106M &  F656N   & 400s\\  \hline
NGC 1818 &U3JS0201M &  F555W  & 1s   \\ 
NGC 1818 &U3JS0202M &  \dots  & 6s   \\ 
NGC 1818 &U3JS0203M &  F160BW & 60s   \\ 
NGC 1818 &U3JS0204M &  \dots &  60s  \\ 
NGC 1818 &U3JS0205M &  \dots & 400s   \\ 
NGC 1818 &U3JS0206M &  \dots & 400s   \\ 
NGC 1818 &U3JS0207M &  \dots & 400s   \\ 
NGC 1818 &U3JS0208M &  F656N  & 350s  \\ \hline
NGC 2004 &U3JS0301M &  F555W  &  1s  \\ 
NGC 2004 &U3JS0302M &  \dots  &  6s  \\ 
NGC 2004 &U3JS0303M &  F160BW &  60s  \\ 
NGC 2004 &U3JS0304M &  \dots &   60s \\ 
NGC 2004 &U3JS0305M &  \dots & 400s   \\ 
NGC 2004 &U3JS0306M &  \dots & 400s   \\ 
NGC 2004 &U3JS0307M &  \dots & 400s   \\ 
NGC 2004 &U3JS0308M &  F656N  &  350s  \\ \hline
NGC 2100 &U3JS0401M &  F555W  & 1s   \\ 
NGC 2100 &U3JS0402M &  \dots  &  6s  \\ 
NGC 2100 &U3JS0403M &  F160BW & 60s   \\ 
NGC 2100 &U3JS0404M &  \dots &  60s  \\ 
NGC 2100 &U3JS0405M &  \dots &  400s  \\ 
NGC 2100 &U3JS0406M &  \dots &  400s  \\ 
NGC 2100 &U3JS0407M &  \dots &  400s  \\ 
NGC 2100 &U3JS0408M &  F656N  &  350s  \\ \hline
\end{tabular}
\end{center}
\caption{The log of observations in the present paper.}
\label{obslog}
\end{table}
 
\clearpage

\begin{table}
\dummytable\label{ngc330phot}
\end{table}
\begin{table}
\dummytable\label{ngc1818phot}
\end{table}
\begin{table}
\dummytable\label{ngc2004phot}
\end{table}
\begin{table}
\dummytable\label{ngc2100phot}
\end{table}

\clearpage
\begin{table}
\begin{center}
\begin{tabular}{ccc} \hline
Locality& A & B \\ \hline
Galaxy  & 5.01& -0.085\\
LMC  & 6.97& -0.153\\
SMC  & 8.02& -0.204\\
 \hline
\end{tabular}
\end{center}
 \caption{Colour excess ratio of E(F160BW$-$F555W) to E($B$$-$$V$) for galactic, 
LMC and SMC reddening laws. Derivied for 10000K $<T_{eff}<$ 25000K. 
E(F160BW$-$F555W) = A$\times$E($B$$-$$V$) + B$\times$E($B$$-$$V$)$\times$(F160BW$-$F555W)}
\label{colourexcess}
\end{table}

\clearpage
\begin{table}
\begin{center}
\begin{tabular}{cc} \hline
Cluster Name & E($B$$-$$V$)  \\ \hline
NGC 330  & 0.08\\
NGC 1818 & 0.08\\
NGC 2004 & 0.08\\
NGC 2100 & 0.26\\ \hline
\end{tabular}
\end{center}
 \caption{Values of E($B$$-$$V$) from the present study.}
\label{reddening}
\end{table}

\begin{table*}
\begin{tabular}[]{l|ccccc}
\hline
& \multicolumn{3}{c|}{F555W $<$ 19} & \multicolumn{1}{c|}{16.5 $<$ F555W $<$ 19} 
& F555W $<$ 16.5
\\
\hline
Cluster
& N$_{\rm MS}$ & N$_{\rm Be}$ & \multicolumn{1}{c|}{N$_{\rm Be}$/N$_{\rm MS}$}& 
\multicolumn{1}{c|}{N$_{\rm Be}$/N$_{\rm MS}$} & N$_{\rm Be}$/N$_{\rm MS}$\\
\hline
NGC 330 &  566 & 100 & 0.18$\pm$0.02 & 0.15$\pm$0.02 & 0.43$\pm$0.10\\
NGC 1818 & 383 & 54 & 0.14$\pm$0.02 & 0.12$\pm$0.02 & 0.27$\pm$0.08\\
NGC 2004 & 415 & 47 & 0.11$\pm$0.02 & 0.08$\pm$0.02 & 0.25$\pm$0.06\\ 
NGC 2100 & 402 & 52 & 0.13$\pm$0.02 & 0.10$\pm$0.02 & 0.26$\pm$0.05\\
\hline
\end{tabular}
\caption{Be star content of clusters in the present study}
\label{betab}
\end{table*}

\begin{table}
\dummytable\label{hsttab}
\end{table}

\clearpage

%
%

%
%

\clearpage

\end{document}